\documentclass[twocolumn,twoside,slac_two]{revtex4}
\usepackage{graphicx}
\usepackage{fancyhdr}
\begin{document}

\title{Precision Electroweak Physics at the Tevatron}


\author{Eric B. James}
\affiliation{FNAL, Batavia, IL 60510, USA \\
on behalf of the CDF and D\O\ Collaborations}

\begin{abstract}

An overview of Tevatron electroweak measurements performed by the 
CDF and D\O\ experiments is presented. The current status and future 
prospects for high precision measurements of electroweak parameters 
and detailed studies of boson production are highlighted. 

\end{abstract}

\maketitle

\thispagestyle{fancy}


\section{Introduction}

The substantial samples of $W$ and $Z$ bosons currently being collected 
by the CDF and D\O\ experiments accommodate a wide variety of precision 
electroweak measurements.  The two general purpose experiments observe 
$p\bar{p}$ collisions at a center-of-mass energy of 1.96 TeV generated 
by the Fermilab Tevatron Collider.  In its current operating mode, the 
Tevatron operates as a $W$ and $Z$ boson factory.  In a normal week of 
operation the Tevatron produces roughly 50,000 $W$ boson and 5,000 $Z$ 
boson events in each lepton decay channel for each experiment.  Currently, 
each of the experiments has recorded approximately 1.5 pb$^{-1}$ of data, 
which corresponds to about a quarter of the total expected Run~II luminosity.

$Z$ boson parameters have been measured to very high precision at the
large electron-positron collider (LEP) at CERN and the linear collider 
at SLAC.  For example, the $Z$ boson mass has been measured with an 
accuracy of 2 parts in 10$^{5}$~\cite{mass}.  However, current 
measurements of the $W$ boson parameters are less precise (the present 
uncertainty on the $W$ boson mass is about 4 parts in 10$^{4}$~\cite{
mass}).  Based on expected Run II integrated luminosities, the 
two Tevatron experiments will collect a sample of $W$ bosons events 
on the same order as the 17 million $Z$ boson events collected by the 
four LEP experiments.  Using these event samples, CDF and D\O\ will 
significantly reduce the current experimental uncertainties on the 
electroweak parameters associated with the $W$ boson.

In addition, the large $W$ and $Z$ boson samples allow for precision 
tests of the QCD production mechanisms for bosons.  In particular, the 
cross section for boson production depends on both the calculable hard 
scattering parton cross sections and the Parton Distribution Functions 
(PDFs), which describe the momentum fractions carried by the quarks 
and gluons within the proton.  The PDFs are determined experimentally, 
and studies of boson production at the Tevatron can be used to place 
constraints on these distributions.  These constraints are important 
because PDF uncertainties significantly impact the level of precision 
of Tevatron measurements of electroweak parameters.  

\section{Detectors}

The CDF and D\O\ detectors are designed to trigger on and accurately 
reconstruct charged particles, electrons, photons, muons, hadronic 
jets, and the transverse energy imbalance associated with neutrinos. 
The $z$-axes of the CDF and D\O\ coordinate systems are defined to be 
along the direction of the incoming protons.  Particle trajectories 
are described by $\theta$, the polar angle relative to the incoming 
proton beam, and $\phi$, the azimuthal angle about the beam axis.  
Pseudorapidity, $\eta = -\mathrm{ln}(\tan(\theta/2)$, is also used 
to describe locations within the detectors. 
  
One particular strength of the CDF detector is its beam-constrained 
central tracking resolution,      
\begin{equation}
\delta(p_{T})/p_{T} \sim 0.0005 \times p_{T}~(\mathrm{GeV}/c)~[|\eta| < 1],
\end{equation}
based on hit information from the outer open-cell drift chamber.  The 
calorimeters of both detectors allow for high-resolution reconstruction 
of the energies of electrons, photons, and jets.  For example, the energy 
resolution for clusters in the CDF central electromagnetic calorimeter is   
\begin{equation}
\delta(E_{T})/E_{T} \sim 13.5\% \oplus 1.5\%~(\mathrm{GeV})~[|\eta| < 1.1],
\end{equation}
which allows for high precision electron energy measurements.  A main 
strength of the D\O\ detector is the forward coverage provided by its 
calorimeters and muon detector systems.  The D\O\ calorimeter provides 
hermetic coverage up to $|\eta| < 4.2$ (compared to $|\eta| < 3.6$ for 
CDF) and muon coverage up to $|\eta| < 2.0$ (compared to $|\eta| < 1.5$ 
for CDF).  This additional forward coverage results in a significantly 
better acceptance for leptons from boson decays, particularly for muons.

\begin{table}[t]
\begin{center}
\caption{Preliminary uncertainty estimates for CDF $W$ 
boson mass measurement using 200~pb$^{-1}$ of data.}
\begin{tabular}{|l|c|c|c|}
\hline \textbf{Uncertainty [MeV]} & \textbf{Electrons} & 
\textbf{Muons} & \textbf{Common}
\\
\hline Lepton Energy Scale & 27 & 17 & 17 \\
       and Resolution & & & \\
\hline Recoil Scale and & 14 & 12 & 12 \\
       Resolution & & & \\
\hline Backgrounds & 7 & 9 & - \\
\hline Production and & 16 & 17 & 16 \\
       Decay Model & & & \\
\hline Statistics & 48 & 53 & - \\
\hline Total & 60 & 60 & 26 \\
\hline
\end{tabular}
\label{tab:wmass}
\end{center}
\end{table}

\section{Measurements of Electroweak Parameters}         

\subsection{$W$ Boson Mass Measurement}

A precision measurement of the $W$ boson mass is among the highest 
priorities for the Tevatron experiments.  Self-energy corrections 
to the $W$ boson depend on the masses of the top quark ($\propto 
M^{2}_{top}$) and the Higgs boson ($\propto \mathrm{ln} M_{H}$), 
as well as potential contributions from non-Standard Model (SM) 
physics.  Because of these dependencies, the $W$ boson mass is a 
critical input to SM fits that constrain the mass of an unobserved 
Higgs boson or, subsequent to a potential Higgs discovery, test the 
consistency of the SM.

The current level of uncertainty on top quark mass measurements 
from the Tevatron experiments~\cite{topmass} is at the level 
of 2.1~GeV/$c^{2}$ which corresponds to roughly a 1.2~$\!\%$
measurement of $M_{top}$.  To obtain equivalent constraining 
power on $M_{H}$, the $W$ boson mass would need to be measured 
to about 0.015~$\!\%$ corresponding to a total uncertainty of 
about 12~MeV/$c^{2}$.  Due to the needed level of precision, 
the $W$ boson mass measurement is extremely challenging.  

In order to make a measurement substantially better than 
0.1~$\!\%$, all aspects of $W$ boson production and 
detection need to be understood at the 10~MeV level.  In 
particular, this precision must be achieved for $W$ boson 
production and decay, lepton momentum/energy scales and 
resolutions, and additional energies within the event 
associated with hadronic recoil against the boson $p_{T}$ 
and underlying interactions of the remnant quarks and gluons.  
Once this detailed event model has been constructed, the $W$ 
boson mass can be determined by generating events for many 
different mass values and picking the set that provides 
the best match with data, in particular by fitting to the 
transverse mass, $M_{T} = E_{T}^{\ell} E_{T}^{\nu} - ( 
E_{x}^{\ell} E_{x}^{\nu} + E_{y}^{\ell} E_{y}^{\nu})$, 
distribution for the $W \rightarrow \ell \nu$ candidate 
events in data.

The CDF experiment is close to completing a $W$ mass measurement 
using 200~pb$^{-1}$ of data collected at the beginning of Run~II.  
The expected uncertainties associated with this measurement are 
shown in Table~\ref{tab:wmass}.  The total uncertainty for the 
combined measurement based on events collected in both the electron 
and muon channels is expected to be 48~MeV/$c^{2}$ which would make 
this measurement the single most precise to date.  More importantly,
the largest component of the total uncertainty is statistical,
indicating that the result will be further improved simply by 
incorporating more data. In fact, with the exception of the 
uncertainty associated with the production and decay model,
each of the uncertainty categories improves with additional 
statistics.  Larger samples of $J/\Psi$, $\Upsilon$, and $Z$ 
boson events, for example, further improve the measurement of 
the track momentum and calorimeter energy scales for leptons.

\begin{figure*}[t]
\centering
\includegraphics[width=135mm]{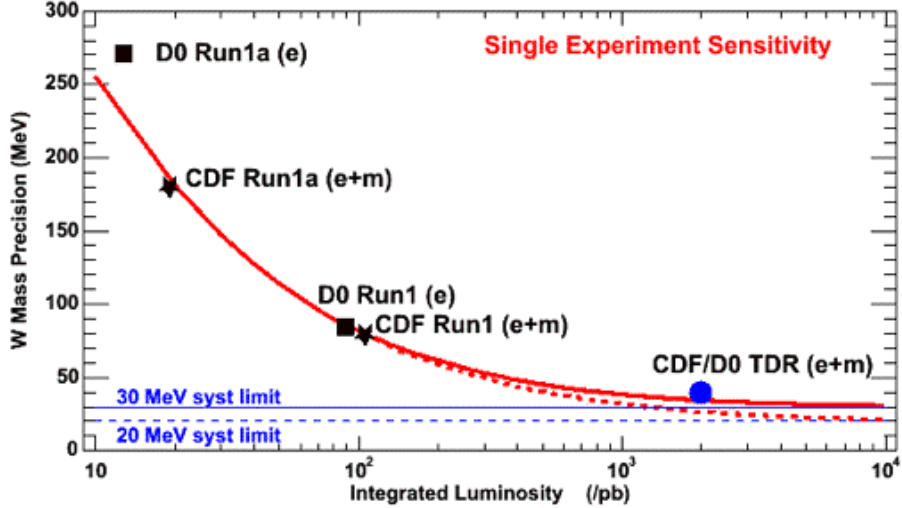}
\caption{Projection for the expected precision of a single 
experiment $W$ mass measurement as a function of integrated 
luminosity based on Run~I Tevatron measurements.} 
\label{fig:wmass}
\end{figure*}
      
\begin{figure}
\includegraphics[width=65mm]{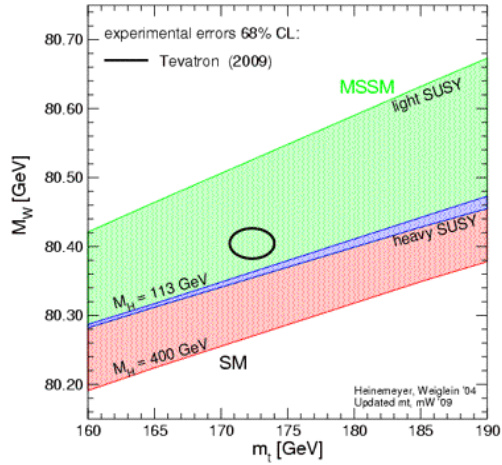}
\caption{Projection for Tevatron constraints on $M_{H}$ 
based on the expected precision of combined top quark 
and $W$ boson mass measurements assuming 8~fb$^{-1}$ of 
data collected by each experiment.}
\label{fig:higgscon}
\end{figure}

Figure~\ref{fig:wmass} shows a projection for the expected 
precision of the $W$ boson mass measurement as a function of 
integrated luminosity for a single experiment based on Tevatron 
Run~I measurements.  The combined preliminary uncertainty for 
the 200~pb$^{-1}$ CDF Run~II analysis lies significantly below 
the expectation based on the Run~I results, indicating improved 
understanding of the $W$ boson event characteristics.  With 
enough additional luminosity, the precision of the measurement
will become limited by the uncertainties associated with the 
boson production and decay model (currently on the order of 
20~MeV) which do not scale with statistics.  Reducing these 
uncertainties requires additional measurements that can
constrain components of the production model, such as PDFs 
and the boson $p_{T}$ spectrum.

A projection for the potential Tevatron constraints on the 
Higgs boson mass based on 8~fb$^{-1}$ of data delivered to 
each experiment is shown in Figure~\ref{fig:higgscon}.  At 
the expected level of precision, significant constraints 
will be placed on non-SM particles such as those predicted 
by supersymmetry (SUSY).

\subsection{$W$ Boson Width Measurements}

The width of the $W$ boson is a less constraining observable 
in global electroweak fits than the mass, but measuring 
its value confirms a basic prediction of the SM and could
provide indications of new physics beyond the SM.  The 
Tevatron experiments make both direct and indirect $W$ boson 
width measurements.  The direct measurements have no built-in 
SM assumptions and are therefore sensitive to potential 
contributions from new physics such as a heavy $W^{\prime}$.  
Indirect measurements are based on SM assumptions and provide 
high precision results that can also be used to place 
constraints on other SM parameters such as CKM matrix elements.        

\begin{figure}
\includegraphics[width=65mm]{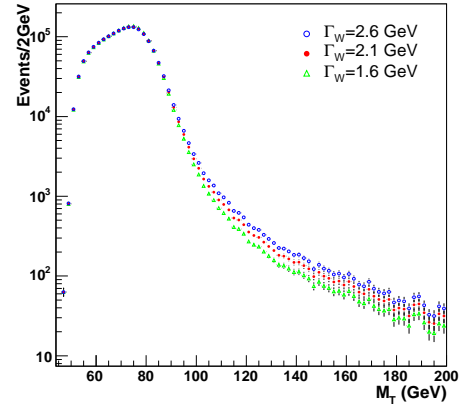}
\caption{D\O\ simulation of the $M_{T}$ distribution for 
$W \rightarrow e \nu$ events as a function of $W$ boson 
width.}
\label{fig:mcwidth}
\end{figure}

\begin{figure}
\includegraphics[width=65mm]{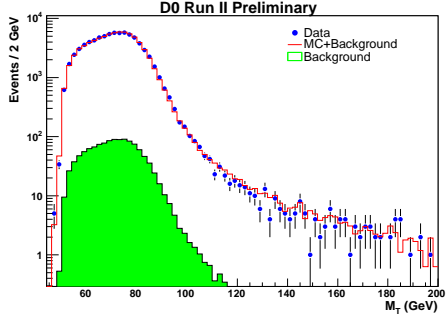}
\caption{D\O\ fit to the $M_{T}$ distribution of 
$W \rightarrow e \nu$ events used to measure the 
$W$ width.}
\label{fig:fitwidth}
\end{figure}

Tevatron direct measurements of the $W$ boson width are 
extracted from the shape of the high $M_{T}$ region in 
$W \rightarrow \ell \nu$ events.  The procedure is similar 
to that used for measuring the $W$ mass.  The $W$ boson 
production and detector resolution effects that distort 
the observed lineshape must be carefully modeled within 
a fast event simulation.  Using the tuned simulation, 
event samples are generated based on a range of input 
values for the $W$ boson width.  The change in the shape 
of the high $M_{T}$ tail as a function of the $W$ width 
is illustrated in Figure~\ref{fig:mcwidth}.  D\O\ has 
made a preliminary direct measurement of the $W$ width 
based on 177~pb$^{-1}$ of Run~II data.  The measurement 
uses the peak region in the $M_{T}$ distribution for $W 
\rightarrow e \nu$ candidate events to normalize signal 
and background contributions to the sample, and then 
fits the shape in the tail region to determine the 
most likely value for the $W$ width.  The final result 
for the $W$ width obtained from the fit shown in 
Figure~\ref{fig:fitwidth} is
\begin{equation}\label{eq:dwidth}
\Gamma_{W} = 2.011 \pm 0.093 (\mathrm{stat}) \pm 0.107 (\mathrm{syst})~\mathrm{GeV} .
\end{equation}
     
Indirect determinations of the $W$ boson width are 
obtained from a measured ratio of production cross 
sections times branching fractions,
\begin{equation}
R = {\sigma \times \mathrm{Br} (W \rightarrow \ell \nu) \over
\sigma \times \mathrm{Br} (Z \rightarrow \ell \ell)} .
\end{equation}
The value of $R$ can be measured very precisely since 
many of the uncertainties associated with the individual
cross section measurements, in particular the significant 
uncertainty on the measured luminosity, cancel in the 
ratio.  Within the context of the SM, this ratio can 
also be expressed as 
\begin{equation}
R = {\sigma(W) \over \sigma(Z)} \times {\Gamma(W \rightarrow \ell \nu) 
\over \Gamma(W)} \times {\Gamma(Z) \over \Gamma(Z \rightarrow \ell \ell)} .
\end{equation}
Using this equation, a precise value for $\Gamma(W)$
can be extracted from $R$ using a next-to-next-to-leading 
order (NNLO) theoretical prediction for $\sigma(W)/\sigma(Z)$, 
precision LEP measurements of $\Gamma(Z \rightarrow \ell \ell)$ 
and $\Gamma(Z)$, and a SM calculation for $\Gamma(W \rightarrow 
\ell \nu)$.

CDF has made an indirect measurement of the $W$ boson 
width based on the first 72~pb$^{-1}$ of data collected 
in Run~II.  The ratio $R$ was measured independently in 
the electron and muon channels, resulting in a combined 
value of     
\begin{equation}
R = 10.84 \pm 0.15 (\mathrm{stat}) \pm 0.14 (\mathrm{syst}),
\end{equation}
which has an overall relative precision of 1.9~$\!\%$.  
Since the most significant contribution to the systematic 
uncertainty on this measurement originates from the lepton 
selection efficiency measurement made from the $Z \rightarrow 
\ell \ell$ data samples, it is expected that a measurement
with a precision of better than 1~$\!\%$ will be possible 
using additional data statistics. 

\begin{figure}
\includegraphics[width=65mm]{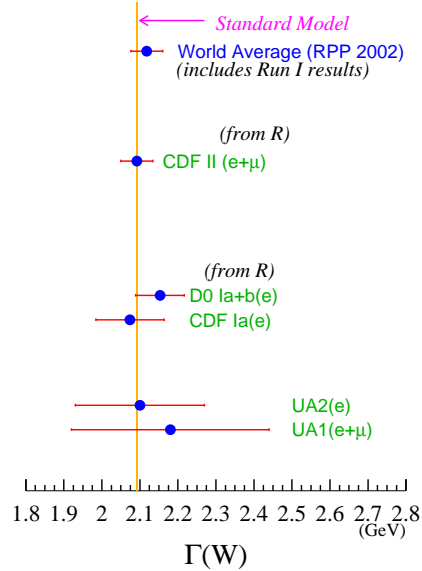}
\caption{Comparison of the CDF indirect width measurement 
with previous results and the SM prediction.}
\label{fig:inwidth}
\end{figure}

The indirect value for the $W$ boson width extracted from
the measured value of $R$ is 
\begin{equation}
\Gamma(W) = 2092 \pm 42~\mathrm{MeV},
\end{equation}
which is in good agreement with the SM prediction and 
the previously described direct measurement of the $W$ 
boson width.  A comparison of the measured indirect width 
with previous results and the SM expectation is shown in 
Figure~\ref{fig:inwidth}.  Since in the SM the total $W$
boson width is a sum over partial widths for leptons and 
quarks, which in the case of the quarks depend on certain 
CKM matrix elements, the measured value of $\Gamma(W)$ can 
also be used to indirectly measure the value of the least 
constrained element, $V_{cs}$.  Based on world-averaged 
measurements of the other CKM matrix elements that 
contribute to the partial widths, CDF obtains a value of            
\begin{equation}
|V_{cs}| = 0.976 \pm 0.030.
\end{equation}
 
\subsection{Quark Couplings}

\begin{figure}
\includegraphics[width=65mm]{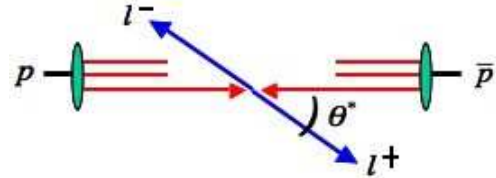}
\caption{Illustration of $\gamma^{\ast}/Z$ decay in the 
parton-parton center of mass frame.  Forward (backward)
events are defined as those with positive (negative)
$cos(\theta^{\ast})$.}
\label{fig:afbpict}
\end{figure}

The Tevatron experiments can extract the axial and vector 
neutral current light quark couplings from measurements of 
the Drell-Yan forward-backward asymmetry.  This asymmetry 
is defined as 
\begin{equation}
A_{FB} = {\sigma_{F} - \sigma_{B} \over \sigma_{F} + \sigma_{B}} 
\end{equation}  
where $\sigma_{F(B)}$ is defined as the cross section for 
Drell-Yan events in which the positively charged lepton is 
produced along (opposite) the proton's direction of motion 
in the parton-parton center of mass frame.  The decay of 
the $\gamma^{\ast}/Z$ in this frame is illustrated in 
Figure~\ref{fig:afbpict}.  The sign of $\cos(\theta^{\ast})$ 
determines whether a given event is forward or backward 
(forward if $\cos(\theta^{\ast}) > 0$).

\begin{figure}[t]
\includegraphics[width=65mm]{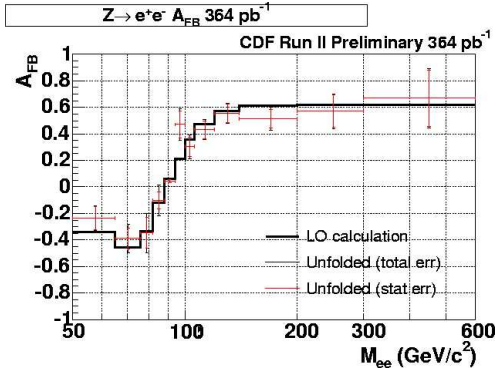}
\caption{CDF Measurement of the forward-backward 
asymmetry in $\gamma^{\ast}/Z \rightarrow e e$ 
events as a function of the di-electron invariant
mass.}
\label{fig:afbres1}
\end{figure}

\begin{figure}[t]
\includegraphics[width=65mm]{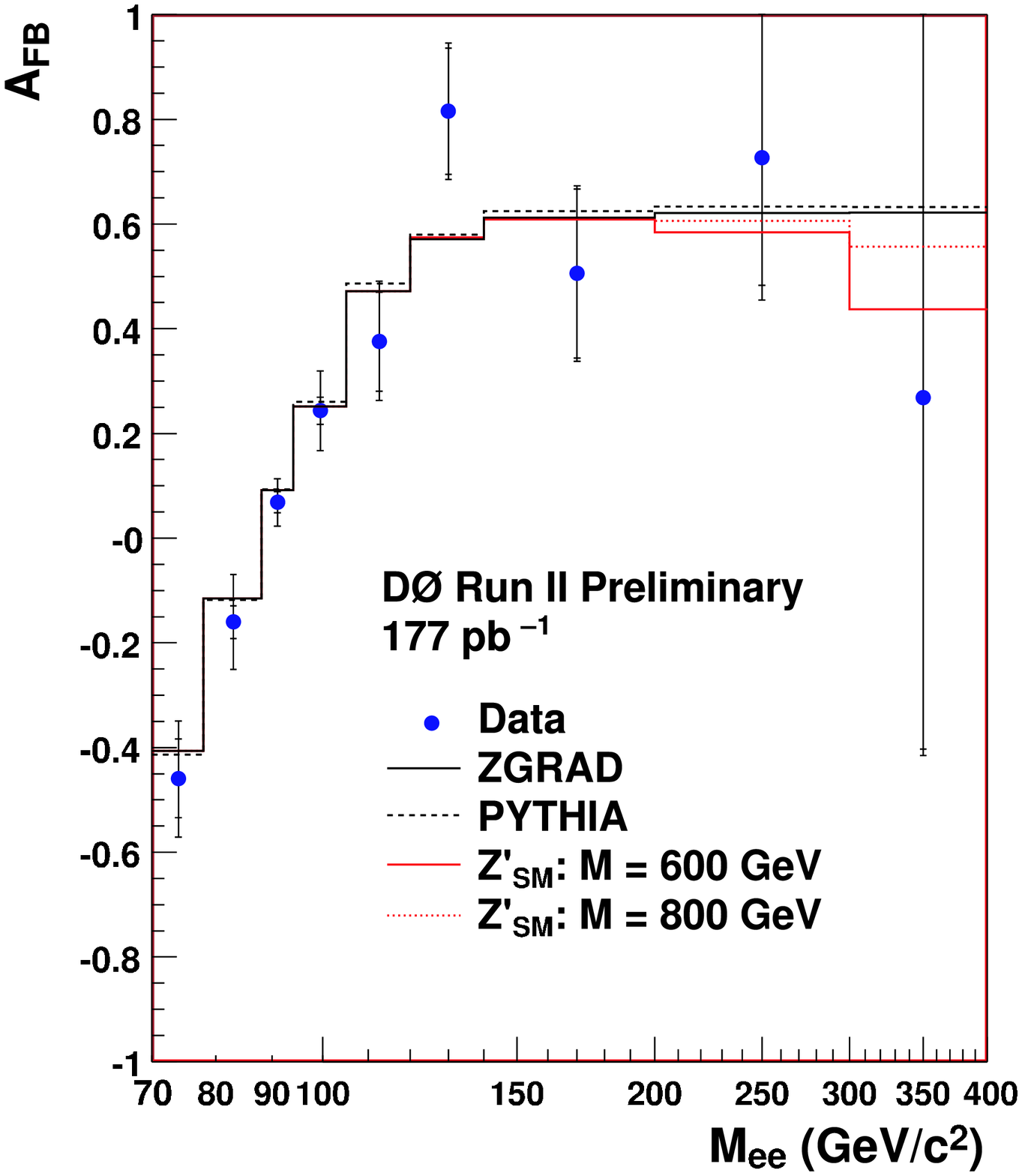}
\caption{D\O\ Measurement of the forward-backward 
asymmetry in $\gamma^{\ast}/Z \rightarrow e e$ 
events as a function of the di-electron invariant
mass.}
\label{fig:afbres2}
\end{figure}

\begin{figure}
\includegraphics[width=65mm]{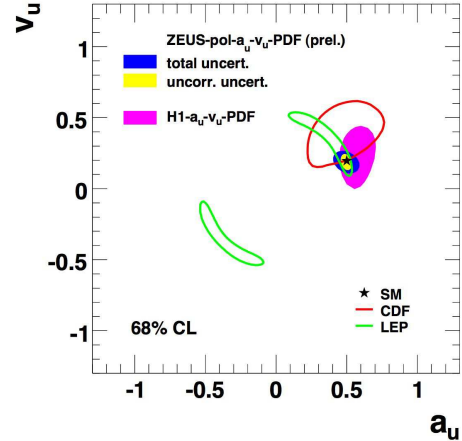}
\caption{Comparison of the limits on the allowed 
range of values for the up quark axial and vector 
neutral current couplings obtained from Tevatron
(72 pb$^{-1}$), HERA, and LEP measurements.}
\label{fig:afbcons}
\end{figure}

CDF and D\O\ have both made preliminary measurements 
of the forward-backward asymmetry for $\gamma^{\ast}/Z 
\rightarrow e e$ events as a function of dielectron 
invariant mass.  The CDF result based on a 364~pb$^{-1}$
data sample is shown in Figure~\ref{fig:afbres1} and 
the D\O\ result based on a 177~pb$^{-1}$ data sample is 
shown in Figure~\ref{fig:afbres2}.  As illustrated in
Figure~\ref{fig:afbcons}, the quark couplings to the $Z$
boson can be extracted from these measurements.  Although
the coupling measurement is less precise than that of the 
LEP experiments, it breaks a two-fold degeneracy in the 
LEP results, providing an important confirmation of the 
SM.  The coupling values have also been determined from 
analysis of deep inelastic scattering data at HERA~\cite{zeus}.

More importantly, the Tevatron experiments measure 
$A_{FB}$ over a wide range of invariant masses (both
below and above the $Z$-pole).  The high mass region
is of particular interest since the effects of new 
bosons interfering with the SM bosons could result 
in measured values of $A_{FB}$ inconsistent with SM 
expectations.  The potential effect of a $Z^{\prime}$ 
on the predicted $A_{FB}$ in the high mass region 
is shown in Figure~\ref{fig:afbres2}, along with 
the measured D\O\ values.  With additional data it 
should be possible to distinguish between the new 
physics and SM scenarios.

\subsection{Trilinear Gauge Couplings}

The analysis of diboson final states at the Tevatron
provides an opportunity for studying self-interactions
of the gauge bosons.  These interactions are a direct 
result of the electroweak SU(2) structure, and the SM
makes specific predictions on the expected production 
cross sections for each diboson final state.  Non-SM
particles that couple to the electroweak bosons can 
modify the expected cross sections, particularly at high 
$E_{T}$, and looking for potential indications of these 
anomalous couplings provides a route to uncovering new 
physics. 

\begin{table}[t]
\begin{center}
\caption{Diboson final states available at the Tevatron
and the trilinear couplings involved in their production.
The couplings shown in parentheses are absent in the SM.}
\begin{tabular}{|l|c|}
\hline \textbf{Diboson Final State} & \textbf{Trilinear Couplings} \\
\hline $q\bar{q}^{\prime} \rightarrow  W \rightarrow W\gamma$ & $WW\gamma$ only \\
\hline $q\bar{q}^{\prime} \rightarrow  W \rightarrow WZ$ & $WWZ$ only \\
\hline $q\bar{q} \rightarrow  \gamma^{\ast}/Z \rightarrow WW$ & $WW\gamma$ , $WWZ$ \\
\hline $q\bar{q} \rightarrow  \gamma^{\ast}/Z \rightarrow Z\gamma$ & $(ZZ\gamma)$ , $(Z\gamma\gamma)$ \\
\hline $q\bar{q} \rightarrow  \gamma^{\ast}/Z \rightarrow ZZ$ & $(ZZ\gamma)$ , $(ZZZ)$ \\
\hline
\end{tabular}
\label{tab:diboson}
\end{center}
\end{table}

Table~\ref{tab:diboson} gives a summary of the diboson 
final states available at the Tevatron and the trilinear 
gauge couplings that contribute to the production of each
state.  The Tevatron experiments are sensitive to different 
combinations of couplings than LEP and explore a higher 
$\sqrt{s}$.  The couplings in the table that are enclosed 
within parentheses are absent in the SM.  Due to the absence 
of these couplings, the associated final states are ideal 
channels for observing effects from new physics.

\begin{figure}
\includegraphics[width=65mm]{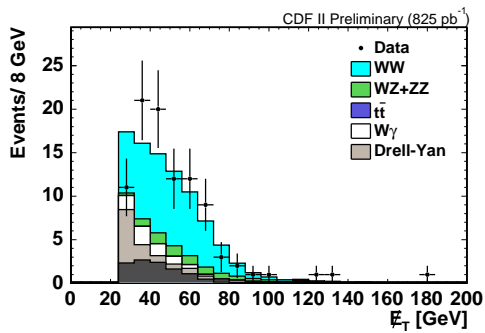}
\caption{Comparison of missing $E_{T}$ distribution for 
observed data events to the combined expectation from
signal and background in the CDF $WW$ analysis.}
\label{fig:WWres}
\end{figure}

The CDF and D\O\ experiments have produced a wide variety 
of new Run~II results based on the study of diboson final 
states~\cite{cdfpub}~\cite{d0pub}.  A few of these are 
highlighted in detail here.  The cross section for $WW$ 
production, which involves both the $WW\gamma$ and $WWZ$ 
trilinear gauge couplings, has recently been measured by 
CDF using a 825~pb$^{-1}$ data sample.  The analysis 
focuses on the dilepton final state produced when both 
$W$ bosons decay into a lepton and neutrino.  Events are 
selected with two opposite-sign leptons (electrons or 
muons) that satisfy the standard CDF selection criteria.  
The missing $E_{T}$ in the event, expected from the 
two neutrinos, is required to be above 25~GeV, greatly 
reducing the main expected background contributions from 
Drell-Yan, $W\gamma$, and $W$ plus jet production.  Before 
looking at the signal region, events in the low missing 
$E_{T}$ region are utilized to cross-check the background
estimation.  In the signal region, the final background 
estimate is $38 \pm 5$ events on top of an expected $WW$ 
signal contribution of $52 \pm 4$ events.  Based on 95 
observed events, CDF measures a cross section of           
\begin{equation}
\sigma(p\bar{p} \rightarrow WW) = 13.6 \pm 3.0 (\mathrm{stat+syst+lum}), 
\end{equation}
consistent with the next-to-leading order (NLO) 
calculation~\cite{ditheory} of $12.4 \pm 0.8$~pb.  The 
final candidate events plotted as a function of event 
missing $E_{T}$, along with the expected signal and 
background contributions, are shown in Figure~\ref{fig:WWres}.

\begin{figure}
\includegraphics[width=65mm]{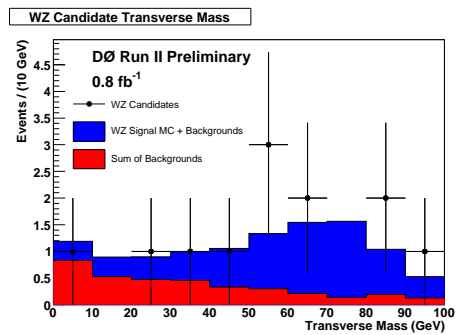}
\caption{Comparison of $M_{T}$ distribution determined from 
the $W$ boson decay products for observed data events with 
the combined expectation from signal and background in the 
D\O\ $WZ$ analysis.}
\label{fig:WZres1}
\end{figure}

Both CDF and D\O\ have also recently completed measurements 
of $WZ$ production.  Production of this final state is of 
particular interest because the $WWZ$ coupling can be studied 
independent of the $WW\gamma$ coupling, which also contributes 
to $WW$ production.  D\O\ has performed a search based on a 
data sample corresponding to roughly 800~pb$^{-1}$ of 
integrated luminosity.  This analysis uses the trilepton 
final state in which both bosons decay leptonically.  A total 
of three leptons (electrons or muons) passing the standard 
D\O\ selection criteria are required.  Of the three leptons, 
two are required to be of the same flavor and form an 
opposite-sign pair with an invariant mass consistent 
with the $Z$ boson mass.  The event missing $E_T$ is also 
required to be greater than 20~GeV, consistent with that 
from the neutrino produced in the decay of the $W$ boson.  
Taking advantage of its wider acceptance for leptons, 
D\O\ expects to see $7.5 \pm 1.2$ signal events on top 
of a background of $3.6 \pm 0.2$ events, and observes 12 
events in the data.  Based on the calculated probability 
for the background to fluctuate into the observed number 
of events, D\O\ obtains 3.3~$\sigma$ evidence for $WZ$ 
production and measures 
\begin{equation}
\sigma(p\bar{p} \rightarrow WZ) = 4.0^{+1.9}_{-1.5} (\mathrm{stat+syst+lum}), 
\end{equation}
consistent with the NLO calculation~\cite{ditheory} of 
$3.68 \pm 0.25$~pb.  Figure~\ref{fig:WZres1} shows the 
transverse mass distribution for the neutrino (missing 
$E_{T}$) and lepton coming from the $W$ boson decay
for the D\O\ candidate events compared to the combined 
expectation from signal and background.

\begin{figure}
\includegraphics[width=65mm]{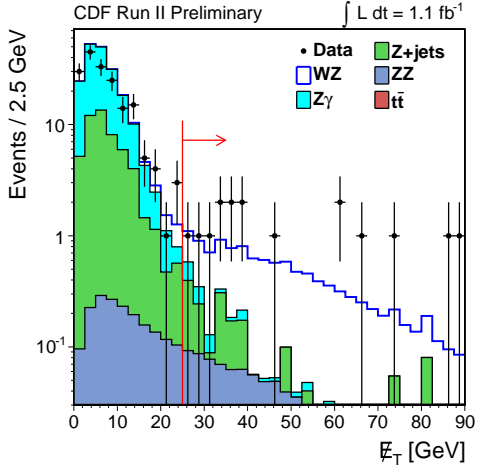}
\caption{Comparison of missing $E_{T}$ distribution for 
observed data events with the combined expectation from
signal and background in the CDF $WZ$ analysis.  The 
arrow on the figure indicates the signal region for this 
search (missing $E_{T} > 25$~GeV).}
\label{fig:WZres2}
\end{figure}

CDF completed a similar search using roughly the same 
amount of data and observed only 2 events compared to
an expectation of $3.7 \pm 0.3$ signal and $0.9 \pm 0.3$
background events.  The observation of two events was 
found to be consistent with both the background-only 
and background plus signal hypotheses.  The smaller 
number of expected events as compared with the D\O\ 
analysis is directly related to the reduced acceptance 
for leptons in the CDF detector.  In order to improve 
the CDF analysis, new lepton categories were created 
to take advantage of additional tracking and calorimeter 
cluster information in the events to increase lepton 
acceptance.  In order to increase electron coverage 
out to $|\eta| < 2.8$, a category for forward electron 
candidates in the calorimeter with no track match was 
added.  Similarly, an increase in muon coverage out to 
$|\eta| < 1.6$ was obtained using forward track candidates 
fiducial to the calorimeter with energy deposits consistent 
with the expectation from a minimum-ionizing particle.  In 
addition, the  tracks pointing at calorimeter cracks were 
placed into a flavor-neutral category of leptons which 
could be assigned as either electrons or muons.  With the 
additional lepton categories in place, CDF performed a new 
search for $WZ$ production using 1.1~fb$^{-1}$ of data.  
Including the improved lepton acceptance, CDF observes 16 
events with signal and background expectations of $12.5 
\pm 0.9$ events and $2.7 \pm 0.4$ events, respectively.  
Based on the probability of the background fluctuating 
into the observed signal, CDF obtains a 5.9~$\sigma$  
observation of $WZ$ production and measures  
\begin{equation}
\sigma(p\bar{p} \rightarrow WZ) = 5.0^{+1.8}_{-1.6} (\mathrm{stat+syst+lum}), 
\end{equation}
which is also consistent with the NLO calculation.
The final candidate events plotted as a function 
of event missing $E_{T}$, along with the expected 
signal and background contributions, are shown in 
Figure~\ref{fig:WZres2}.

\begin{figure}
\includegraphics[width=65mm]{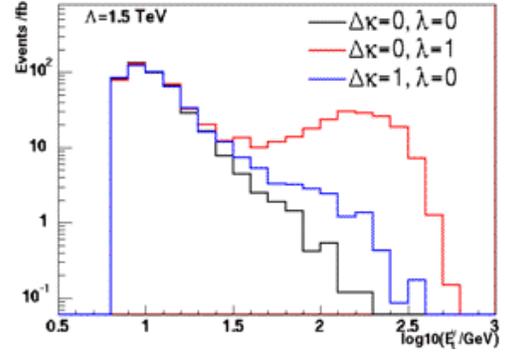}
\caption{The predicted shape of a generic diboson cross 
section as a function of $E_{T}$ for different values of 
$\lambda$ and $\Delta \kappa$.}
\label{fig:acpict}
\end{figure}

As mentioned previously, diboson production is sensitive 
to new physics appearing in the trilinear gauge couplings.
Potential new physics contributions can be incorporated in 
the Lagrangian using a standard methodology that introduces
two parameters, $\lambda$ and $\Delta\kappa$, which are zero 
in the SM and non-zero in the case of additional new physics 
contributions.  Generally, the effect of anomalous couplings 
on diboson production is a net increase in the cross section 
at high $E_{T}$.  Figure~\ref{fig:acpict} illustrates how the 
shape of the diboson cross section as a function of $E_{T}$ 
varies for different values of $\lambda$ and $\Delta \kappa$.      
The added terms in the Lagrangian violate unitarity unless 
an upper limit ($\Lambda$) on the scale for the new physics 
is imposed.  A common approach is to use the parameterization
$\alpha(s) = \alpha_{0}/(1 + s/\Lambda^{2})^{2}$ which causes 
the effect of the anomalous couplings to ``turn-off'' as the 
upper limit on the energy scale is approached.  

\begin{figure}
\includegraphics[width=65mm]{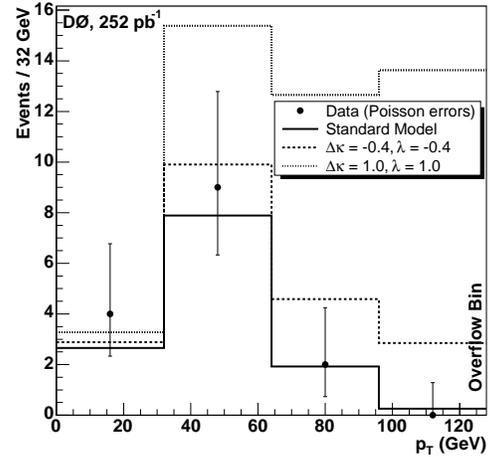}
\caption{Comparison of leading lepton $p_{T}$ distribution
for D\O\ $WW$ candidate events observed in the dilepton 
final state with SM and non-SM expectations.}
\label{fig:acdata}
\end{figure}

\begin{figure*}[t]
\centering
\includegraphics[width=135mm]{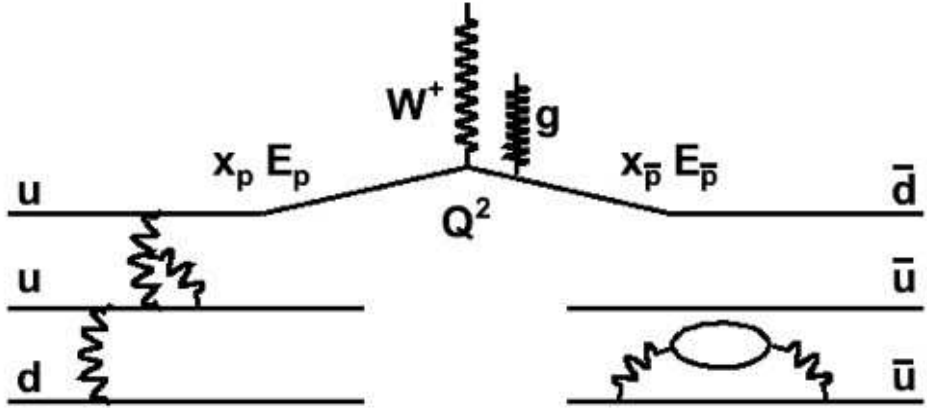}
\caption{Illustration of $W$ boson production at the Tevatron.
A $u$ quark in the proton annihilates with a $\bar{d}$ quark 
in the anti-proton at a squared center of mass $s = Q^{2}$ to 
produce a $W^{+}$.  The energies of the $u$ and $\bar{d}$ are 
$x_{p}E_{p}$ and $x_{\bar{p}}E_{\bar{p}}$, respectively.} 
\label{fig:bosonprd}
\end{figure*}     
   
D\O\ has performed a preliminary analysis to set anomalous
couplings limits based on a measurement of the $WW$ cross 
section using dilepton final states.  The analysis sets 
limits on anomalous $WW\gamma$ and $WWZ$ trilinear gauge
couplings under the assumption that the two couplings 
are equal and $\Lambda = 2$~TeV.  Figure~\ref{fig:acdata}    
shows the D\O\ data and both SM and non-SM expectations 
plotted as a function of the $p_{T}$ of the highest $p_{T}$ 
lepton.  Based on the observed agreement between data and 
the SM prediction, D\O\ obtains the following limits:
\begin{equation}
-0.32 < \Delta \kappa < 0.45 , -0.29 < \lambda < 0.45 . 
\end{equation}
These preliminary limits can be improved significantly with
larger data samples and incorporating information from other 
final states.

\section{Studies of Boson Production}

\subsection{Boson Production at the Tevatron}

A typical example of boson production at the Tevatron is 
shown in Figure~\ref{fig:bosonprd}.  At leading order (LO), 
a quark and anti-quark pair annihilate to create a $W$ or 
$Z$ boson, which subsequently decays into a quark or lepton
pair.  The production cross section is calculated as a sum 
of partial cross sections ($d\sigma_{q\bar{q}}$), convoluted 
with the PDFs that describe the distributions of the proton 
momentum fraction ($x_{p}$) carried by each of the constituent 
quarks and gluons.  The cross section can be written as   
\begin{equation}
\sigma = \int \sum_{i,j}[f_{i}^{q}(x_{p})f_{j}^{\bar{q}}(x_{\bar{p}}) +  
f_{i}^{\bar{q}}(x_{p})f_{j}^{q}(x_{\bar{p}})] \times d\sigma_{q\bar{q}} 
dx_{p} dx_{\bar{p}}
\end{equation}
where $i$ and $j$ denote the different possible quark 
flavor combinations.  The longitudinal momentum of 
the produced boson is directly related to the PDFs.  
In particular, if one of the two annihilating quarks 
carries a significantly larger fraction of proton 
momentum, the boson will be produced with momentum 
in the same direction as the incident proton.

The effects of QCD and QED NLO corrections are also 
important.  QCD corrections give rise to final states 
that contain multiple partons, sometimes with high $p_{T}$, 
and modify the overall boson production kinematics, including 
the boson $p_{T}$ spectrum.  The most important effect 
originating from NLO QED corrections is photon radiation 
from final state charged leptons, which have a significant 
effect on lepton identification and kinematics.  QED 
radiation from the initial state quarks and from the boson 
itself (in the case of $W$ bosons) also contributes to the 
overall event kinematics. 
    
\subsection{Parton Distribution Functions}

The functional forms of the PDFs originate from 
non-perturbative QCD interactions and are therefore 
incalculable.  Instead, they are parameterized using 
data from deep inelastic scattering, fixed target, 
and hadron collider experiments.  Two standard 
parameterizations come from the CTEQ~\cite{cteq} and 
MRST~\cite{mrst} groups.  In the case of the CTEQ 
group, the parton momentum fraction distributions 
are parameterized as  
\begin{equation}
xf_{a}(x,Q_{0}) = A_{0}x^{A_{1}}(1-x)^{A_{2}}e^{A_{3}x}(1+A_{4}x)^{A_{5}} 
\end{equation}
for five categories of quark/gluon proton constituents
(valence $u$ and $d$ quarks, sea $\bar{u}$ and $\bar{d}$ 
quark combinations, and gluons).  This configuration 
gives a total of thirty free parameters in the fit to 
the experimental data, although the CTEQ group chooses 
to leave ten of these at fixed values.  The remaining free
parameters are determined for a low energy scale, $Q_{0} =
1.3$~GeV, and the $Q^{2}$ dependence is obtained from QCD
evolution equations.

\begin{figure*}[t]
\centering
\includegraphics[width=135mm]{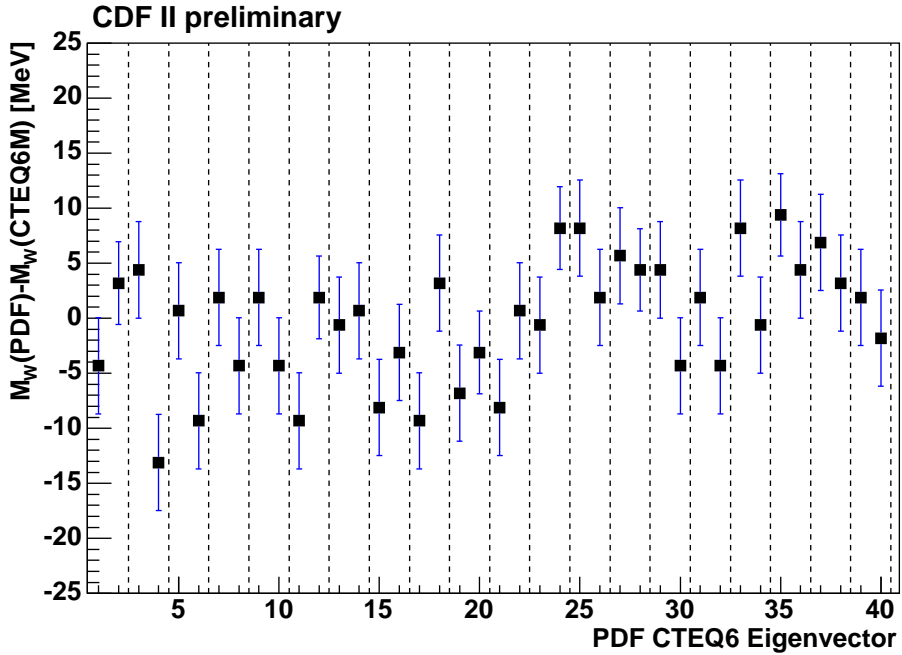}
\caption{An example of how CTEQ ``error'' PDF sets are 
used to determine an overall PDF uncertainty.  The shift 
in the measured $W$ boson mass from its central value is 
obtained for Monte Carlo templates generated with each 
of the 40 error PDF sets.  The observed shifts associated 
with each of the twenty orthogonal eigenvectors are added 
in quadrature to determine the total uncertainty.} 
\label{fig:wpdf}
\end{figure*}

A recent development is that each group also provides a set 
of ``error'' PDFs that are intended to map out the allowable 
parameter space for the PDFs within the experimental data 
uncertainties.  The twenty free parameters used in the fit 
are found to be correlated with one another.  To facilitate 
uncertainty calculations, these correlations are removed by 
forming eigenvectors within the $A_{i}$-space.  For each of 
the twenty eigenvectors, two complete PDF sets are generated 
corresponding to a given increase in $\chi^{2}$ of the overall 
fit ($\Delta \chi^{2} = 100$ for the CTEQ group).  The MRST 
group follows a similar procedure using a slightly different 
parameterization that results in only fifteen free parameters 
for their fit.  The MRST group also uses a smaller $\Delta 
\chi^{2} = 50$ to construct its version of the error PDFs.

An example of how error PDFs are used to determine an 
overall PDF uncertainty for a specific analysis is shown 
in Figure~\ref{fig:wpdf} for the case of the $W$ boson mass 
measurement.  The shift in the measured mass from its central 
value is obtained using Monte Carlo templates generated with 
each of the forty error PDF sets.  Since the twenty eigenvectors 
are orthogonal to each other by design, the observed shifts 
associated with each can be added in quadrature to determine a 
total PDF model uncertainty.  Although each eigenvector typically 
contains information about multiple fit parameters, there is a 
strong correlation in some cases between a given fit parameter 
and an eigenfunction.  For example, the eigenvector corresponding 
to error PDFs 1 and 2 in Figure~\ref{fig:wpdf} has a significant 
correlation with the $A_{1}$ (low-$x_{p}$) parameter associated 
with valence $u$ quarks.  These correlations give an indication 
of the experimental inputs to the fits which need to be improved 
to reduce the overall PDF uncertainty for a specific analysis.

\subsection{Inclusive Production Cross Sections}

\begin{figure}
\includegraphics[width=65mm]{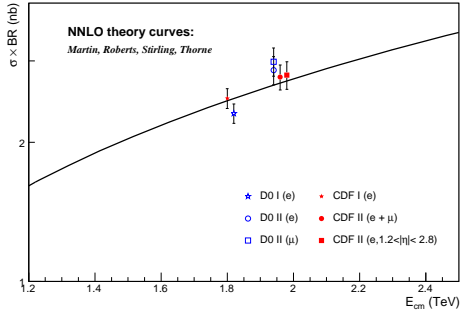}
\caption{Summary of Tevatron inclusive $W$ boson cross section 
measurements as a function of $E_{CM}$ compared to a NNLO
theoretical calculation (solid black line).}
\label{fig:wxsec}
\end{figure}

\begin{figure}
\includegraphics[width=65mm]{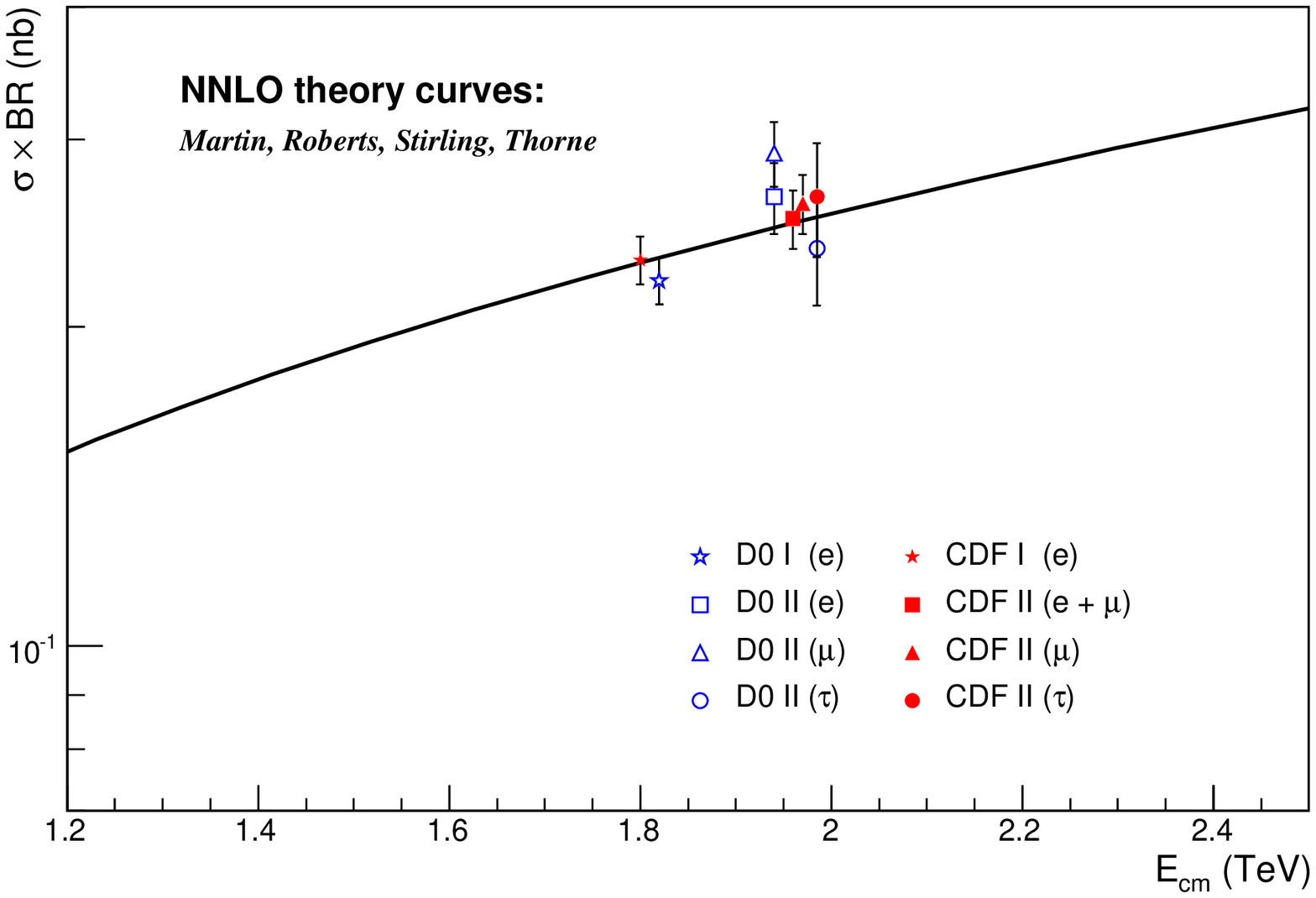}
\caption{Summary of Tevatron inclusive $Z$ boson cross section 
measurements as a function of $E_{CM}$ compared to a NNLO
theoretical calculation (solid black line).}
\label{fig:zxsec}
\end{figure}

Because many electroweak measurements at the Tevatron are 
sensitive to uncertainties in the PDF model, both CDF and 
D\O\ perform studies of boson production to constrain the 
PDF model. The simplest of these studies are measurements 
of the inclusive boson production cross sections.  The 
Tevatron experiments measure inclusive $W$ and $Z$ cross 
sections using each of the lepton ($e$, $\mu$, and $\tau$) 
decay channels.  The dominant uncertainty in these results 
is associated with the integrated luminosity measurements 
made by each experiment ($\sim 6\%$).  Within this uncertainty, 
the measured cross sections are found to be in good agreement 
with the NNLO theoretical calculations~\cite{inclusive}.  
The agreement between CDF and D\O\ measured values and 
theoretical predications are shown in Figures~\ref{fig:wxsec} 
and~\ref{fig:zxsec}.  Since the theoretical uncertainties are 
significantly smaller than the measurement uncertainties, no 
additional constraints on the boson production model can be 
obtained from these measurements.   

\subsection{Forward $W$ Boson Cross Section}

\begin{figure}
\includegraphics[width=65mm]{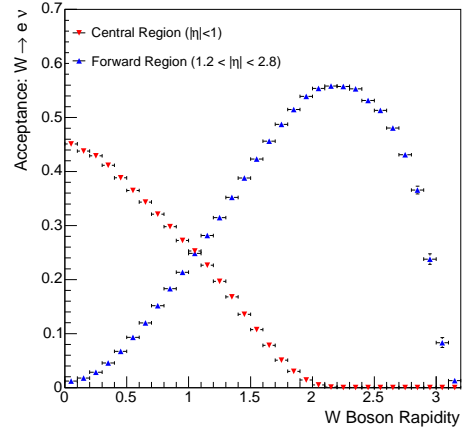}
\caption{$W$ boson acceptance as a function of rapidity for 
the CDF $W \rightarrow e \nu$ cross section measurements 
using events with reconstructed electrons in the central
and forward parts of the detector.}
\label{fig:bosrap}
\end{figure}

Differential cross section measurements contain additional 
information that can be used to constrain PDFs.  CDF 
performs a simple differential measurement by independently
evaluating the $W$ boson cross section using $W \rightarrow
e \nu$ events with electrons observed in the central and 
forward regions of the detector.  Figure~\ref{fig:bosrap} 
shows the $W$ boson acceptance as a function of the boson
rapidity, defined as
\begin{equation}
y_{W} = {1 \over 2} \mathrm{log} {E + p_{z} \over E - p_{z}},  
\end{equation}
for the CDF $W \rightarrow e \nu$ cross section measurements
using events with electrons reconstructed in the central 
and forward calorimeter modules.  Since $W$ bosons produced 
at different rapidities probe different regions of $x_{p}$, 
the ratio of central to forward cross sections measurements
can be a useful tool for placing constraints on PDFs.

\begin{figure}
\includegraphics[width=65mm]{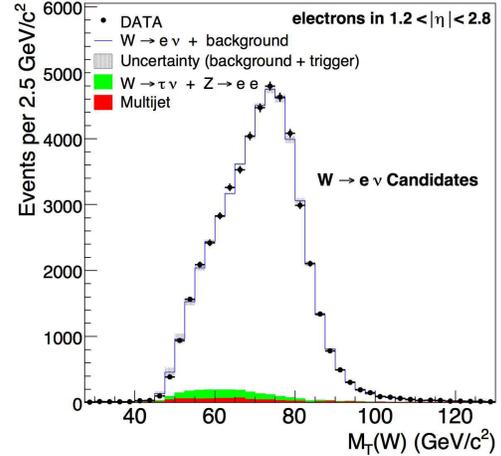}
\caption{$M_{T}$ distribution for candidate events in the 
CDF cross section measurement based on $W \rightarrow e 
\nu$ events with electrons in the forward detector region.}
\label{fig:fxsec}
\end{figure}

The selection of forward electron candidates is based 
on electromagnetic clusters in the calorimeter matched 
with tracks reconstructed primarily from silicon detector 
hits~\cite{ionim}.  Given the selection criteria, CDF 
observes 48,165 candidate events in a 223~pb$^{-1}$ data 
sample.  The $M_{T}$ spectrum of the candidate events is 
shown in Figure~\ref{fig:fxsec}, along with the combined 
expectation for signal and background.  The observed 
agreement indicates a good understanding of the forward 
detector systems.

The measured forward ($1.2~<~|\eta^{det}_{e}|~<~2.8$) cross 
section is 
\begin{equation}
\sigma^{for} = 2796 \pm 13 (\mathrm{stat}) ^{+95}_{-90} (\mathrm{syst})~\mathrm{pb},
\end{equation}
neglecting the luminosity uncertainty which will cancel in 
the cross section ratio.  The previously measured central 
($|\eta^{det}_{e}|~<~0.9$) cross section~\cite{ourprl} has 
a value of  
\begin{equation}
\sigma^{cen} = 2771 \pm 14 (\mathrm{stat}) ^{+62}_{-56} (\mathrm{syst})~\mathrm{pb},
\end{equation}
also neglecting the luminosity uncertainty.  The remaining 
systematic uncertainties on the measurements are dominated 
by those associated with electron identification and the 
PDF model.  In order to separate these, CDF uses visible 
cross sections, defined as    
\begin{equation}
\sigma_{vis} = \sigma_{tot} \times A ,
\end{equation}
where $A^{cen}$ is for example the kinematic and geometric 
acceptance for $W \rightarrow e \nu$ events in the central 
cross section measurement.  Using this definition the PDF 
model uncertainties are removed from the measured ratio of 
cross sections,
\begin{equation}
R_{exp} = \sigma^{cen}_{vis} / \sigma^{for}_{vis} = 0.925 \pm 0.033 .
\end{equation} 
CDF then compares the measured ratio with the equivalent 
theoretical ratio of acceptances
\begin{equation}
R_{th} = A^{cen} / A^{for}
\end{equation}
determined from simulated event samples generated using both 
the CTEQ ($R_{th} = 0.924 \pm 0.037$) and MRST ($R_{th} = 
0.941 \pm 0.012$) PDF distributions.  The uncertainties on 
the acceptance ratios are obtained from the error PDF sets 
using the previously described method.  The uncertainty 
on the measured ratio is of the same order as the PDF 
uncertainties on the theoretical ratio, suggesting that a 
similar measurement with additional statistics would help 
to constrain the PDF models.  

\subsection{Differential $Z$ Boson Cross Section}

\begin{figure}
\includegraphics[width=65mm]{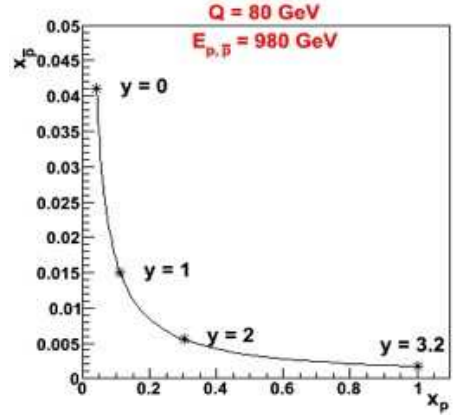}
\caption{The interacting partons' momentum fractions 
required to produce a $W$ boson  ($Q = 80$~GeV).  The 
larger the difference between $x_{p}$ and $x_{\bar{p}}$,
the greater the rapidity of the produced boson.}
\label{fig:yrap}
\end{figure}

Measuring the differential boson production cross section 
over the full rapidity range can further improve PDF model 
constraints.  The dilepton decay modes of the $Z$ boson 
allow for precise measurements, since the backgrounds in 
these final states are small and the full event kinematics 
can be precisely reconstructed.  The rapidity of the $Z$ 
boson is closely related to the proton momentum fractions 
carried by the two colliding quarks.  As shown in 
Figure~\ref{fig:yrap}, $W$ or $Z$ bosons are produced at 
high rapidity when the proton momentum fraction of one 
quark is significantly larger than that of the other.  
Therefore, the measured differential cross section at high 
rapidity is a good probe of the PDF distributions at high 
$x_{p}$.

\begin{figure}
\includegraphics[width=65mm]{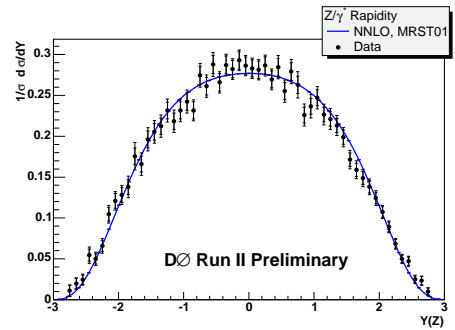}
\caption{Differential $Z$ boson cross section measured by 
D\O\ as a function of boson rapidity. The measured cross 
section is in good agreement with a NNLO theoretical 
prediction based on MRST PDFs (solid line).}
\label{fig:yres}
\end{figure}

D\O\ has made a preliminary measurement of the differential
$Z$ boson cross section based on a 337~pb$^{-1}$ data sample.
Using $Z \rightarrow e e$ candidate events, D\O\ reconstructs
the differential cross section shown in Figure~\ref{fig:yres}.   
The measured cross section is observed to agree well with the 
NNLO prediction.  The measurement is currently statistics-limited 
but can be used to constrain PDF models using additional data.  

\subsection{$W$ Boson Charge Asymmetry}

\begin{figure}
\includegraphics[width=65mm]{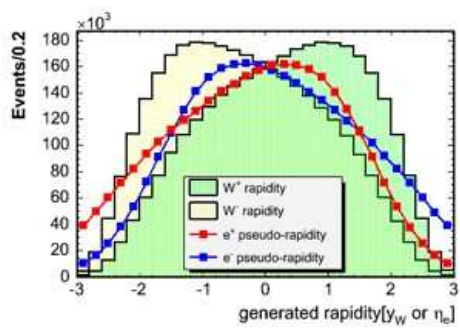}
\caption{Rapidity distributions of positively and 
negatively charged $W$ bosons produced at the 
Tevatron, and the pseudorapidity distributions 
of the positively and negatively charged leptons 
produced in their decays.}
\label{fig:wasym}
\end{figure}

A final measurement useful for constraining PDFs is the 
$W$ boson charge asymmetry measurement.  On average the 
$u$ quarks inside the proton contain a higher fraction 
of the proton's momentum than the $d$ quarks.  Due to 
this imbalance, $W^{+}$ ($W^{-}$) bosons produced at the 
Tevatron have a net positive (negative) rapidity, as 
shown in Figure~\ref{fig:wasym}.  The V-A structure of the 
electroweak couplings dictates the angular distribution 
of the leptons in the decays of the $W$ bosons, which is 
preferentially opposite to the production asymmetry.  As 
shown in Figure~\ref{fig:wasym}, the net effect of the 
decay asymmetry is to partially reduce the observable 
production asymmetry extracted from the lepton rapidity 
distributions.  Because the production asymmetry originates 
from the imbalance of the momentum fractions carried by 
$u$ and $d$ quarks within the proton, charge asymmetry 
measurements provide constraints on the $d/u$ ratio in 
the proton as a function of $x_{p}$.

\begin{figure}
\includegraphics[width=65mm]{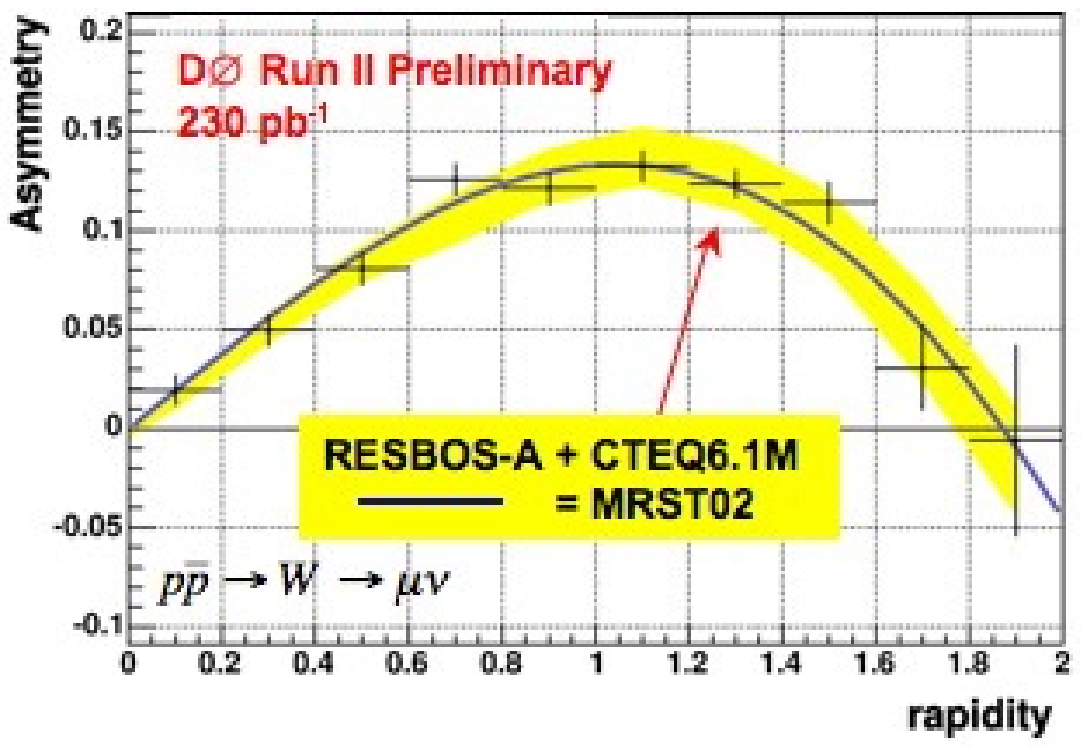}
\caption{D\O\ lepton charge asymmetry measurement based 
on $W \rightarrow \mu \nu$ events.  The measurement is 
compared to a theoretical calculation based on the 
CTEQ and MRST PDF models.}
\label{fig:d0asym}
\end{figure}

\begin{figure}
\includegraphics[width=65mm]{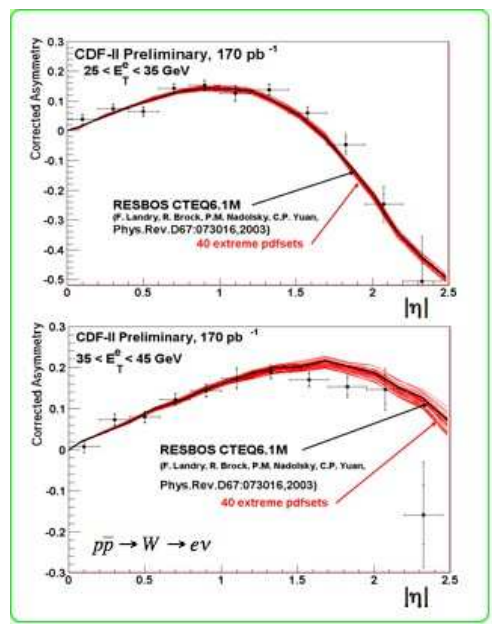}
\caption{CDF lepton charge asymmetry measurement based 
on $W \rightarrow e \nu$ events.  The measurement is 
compared to a theoretical calculation based on the 
CTEQ PDF model.}
\label{fig:cdfasym}
\end{figure}

Measurements are typically performed using the charged
leptons from the $W$ boson decays.  The lepton asymmetry
is defined as 
\begin{equation}
A(\eta_{\ell}) = {d\sigma_{+}/d\eta_{\ell} - d\sigma_{-}/d\eta_{\ell} \over
d\sigma_{+}/d\eta_{\ell} + d\sigma_{-}/d\eta_{\ell}} = A(y_{W}) \otimes (V-A) .
\end{equation}
Both CDF and D\O\ have performed preliminary measurements
of the lepton charge asymmetry.  The key experimental 
issues are understanding forward lepton identification 
and charge misidentification rates, which are needed to 
correct the observed asymmetry.  A D\O\ measurement of 
the lepton charge asymmetry using $W \rightarrow \mu 
\nu$ events selected from a 230~pb$^{-1}$ data sample is 
shown in Figure~\ref{fig:d0asym}.  The measured charge 
misidentification rates for this analysis are found to 
be below $10^{-4}$ out to muon pseudorapidities of $2$.  
The measured asymmetry is compared to a theoretical 
prediction based on the CTEQ PDF model.  The measurement 
is observed to have some sensitivity to PDFs even at 
the current level of statistical sensitivity.  The CDF 
measurement based on $W \rightarrow e \nu$ events 
selected from a 170~pb$^{-1}$ data sample are shown in 
Figure~\ref{fig:cdfasym}.  Here the data is separated 
into two categories based on the $E_{T}$ of the electron.  
Comparisons with theoretical predictions using the CTEQ 
PDF model illustrate the increased sensitivity of the high 
$E_{T}$ events to PDF variations.

\begin{figure}
\includegraphics[width=65mm]{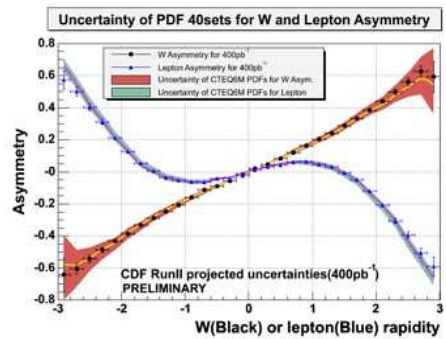}
\caption{Comparison of the potential PDF sensitivities for 
lepton charge asymmetry and $W$ boson production asymmetry 
measurements made with a common set of simulated candidate 
events corresponding to a luminosity of 400~pb$^{-1}$.}
\label{fig:newasym}
\end{figure}

A new generation of Tevatron charge asymmetry analyses
are currently under development, with the goal of fully 
exploiting the kinematic information in $W$ events to 
directly reconstruct the underlying $W$ boson production 
asymmetry.  Applying a $W$ mass constraint leads to two 
kinematic solutions that can be weighted by taking into 
account information about the production and decay of the 
$W$ bosons.  Potential dependencies on the input model are 
resolved through an iterative procedure.  Preliminary CDF 
studies of this approach indicate significantly increased 
sensitivity to PDFs.  The potential increase in sensitivity 
is illustrated in Figure~\ref{fig:newasym}, which shows 
a comparison of hypothetical lepton charge asymmetry and 
direct $W$ boson charge asymmetry measurements based on a 
common set of candidate events obtained from a 400~pb$^{-1}$ 
dataset.
     
\section{Conclusions}

The large samples of $W$ and $Z$ bosons being collected 
at the Tevatron accommodate a wide variety of electroweak 
measurements.  In particular, the properties of the $W$ 
boson can be measured with very high precision by the CDF 
and D\O\ experiments.  In addition, detailed studies of 
boson production at the Tevatron can be used to constrain 
PDF models and provide important information about the 
boson production mechanisms.  The analyses reported here 
are based on only a small fraction of the expected data, 
so there is significant room for improving the precision 
of the current measurements.  It is important to note that 
obtaining similar precision results from the Large Hadron 
Collider (LHC) will be challenging and certainly require 
input (such as PDF constraints) from the Tevatron 
experiments. 

\bigskip % extra skip inserted

\begin{thebibliography}{0}
\expandafter\ifx\csname natexlab\endcsname\relax\def\natexlab#1{#1}\fi
\expandafter\ifx\csname bibnamefont\endcsname\relax
  \def\bibnamefont#1{#1}\fi
\expandafter\ifx\csname bibfnamefont\endcsname\relax
  \def\bibfnamefont#1{#1}\fi
\expandafter\ifx\csname citenamefont\endcsname\relax
  \def\citenamefont#1{#1}\fi
\expandafter\ifx\csname url\endcsname\relax
  \def\url#1{\texttt{#1}}\fi
\expandafter\ifx\csname urlprefix\endcsname\relax\def\urlprefix{URL }\fi
\providecommand{\bibinfo}[2]{#2}
\providecommand{\eprint}[2][]{\url{#2}}

\end{thebibliography}


\begin{thebibliography}{9}   % Use for  1-9  references
%\begin{thebibliography}{99} % Use for 10-99 references

\bibitem{mass}
W.-M. Yao {\it et al.}, J. Phys. G{\bf33}, 1 (2006), and
references therein.
\bibitem{topmass}
Tevatron Electroweak Working Group (for the CDF and D\O\
Collaborations), ``Combination of CDF and D\O\ results on
the Mass of the Top Quark,'' hep-ex/060832.
\bibitem{zeus}
ZEUS Collaboration, ``QCD and Electroweak analysis of 
the ZEUS NC and CC inclusive and jet cross sections,'' 
http://www-zeus.desy.de/public\_results/publicsearch.html,
ZEUS-prel-06-003.
\bibitem{cdfpub}
http://www-cdf.fnal.gov/physics/ewk/
\bibitem{d0pub} 
http://www-d0.fnal.gov/Run2Physics/wz/
\bibitem{ditheory}
J. Campbell and K. Ellis, Phys. Rev. D{\bf60}, 113006 (1999).
\bibitem{cteq}
http://www.phys.psu.edu/cteq 
\bibitem{mrst}
http://www-spires.dur.ac.uk/hepdata/mrs.html 
\bibitem{inclusive}
P. Sutton, A. Martin, R. Roberts, and W. Stirling, 
Phys. Rev. D{\bf45}, 2349 (1992).
\bibitem{ionim}
C. Hays {\it et al.}, 
Nucl. Instrum. Meth. Phys. Res. A {\bf 538}, 249 (2005).
\bibitem{ourprl}
D. Acosta {\it et al.},
Phys. Rev. Lett. {\bf 94}, 091803 (2005).

\end{thebibliography}

\end{document}